\documentclass{article}

\PassOptionsToPackage{numbers, compress}{natbib}

\usepackage[preprint]{neurips_2026}

\usepackage[utf8]{inputenc} 
\usepackage[T1]{fontenc}    
\usepackage{hyperref}       
\usepackage{url}            
\usepackage{booktabs}       
\usepackage{amsfonts}       
\usepackage{nicefrac}       
\usepackage{microtype}      
\usepackage{xcolor}         
\usepackage[most]{tcolorbox}
\usepackage{amsmath}
\usepackage{stmaryrd}
\usepackage{cleveref}

\usepackage[utf8]{inputenc} 
\usepackage[LGR, T1]{fontenc}
\usepackage{makecell}
\usepackage[framemethod=TikZ]{mdframed}
\usepackage{tikz}
\usetikzlibrary{positioning, arrows.meta, backgrounds, fit}
\usepackage{amssymb}
\usepackage{pifont}

\usepackage{enumitem}
\usepackage{xspace}

\usepackage{minted}
\usepackage{wrapfig}
\usepackage{clrscode}
\usepackage{subcaption}
\usepackage{inconsolata}
\usepackage{tikz-cd}
\usepackage{graphicx}
\usepackage{adjustbox}

\usepackage{color}
\definecolor{keywordcolor}{rgb}{0.7, 0.1, 0.1}   
\definecolor{tacticcolor}{rgb}{0.0, 0.1, 0.6}    
\definecolor{commentcolor}{rgb}{0.4, 0.4, 0.4}   
\definecolor{symbolcolor}{rgb}{0.0, 0.1, 0.6}    
\definecolor{sortcolor}{rgb}{0.1, 0.5, 0.1}      
\definecolor{attributecolor}{rgb}{0.7, 0.1, 0.1} 

\usepackage{textcomp}
\usepackage[T1]{fontenc}
\usepackage[utf8]{inputenc} 
\usepackage{xcolor}
\usepackage{amssymb}
\usepackage{listings}

\definecolor{keywordcolor}{rgb}{0.7, 0.1, 0.1}
\definecolor{tacticcolor}{rgb}{0.0, 0.1, 0.6}
\definecolor{commentcolor}{rgb}{0.4, 0.4, 0.4}
\definecolor{symbolcolor}{rgb}{0.0, 0.1, 0.6}
\definecolor{sortcolor}{rgb}{0.1, 0.5, 0.1}
\definecolor{attributecolor}{rgb}{0.7, 0.1, 0.1}

\definecolor{promptbg}{rgb}{0.96,0.96,0.96}
\definecolor{promptframe}{rgb}{0.75,0.75,0.75}

\lstdefinestyle{leanplain}{
    language=lean,
    basicstyle=\footnotesize\ttfamily\color{black},
    backgroundcolor=\color{promptbg},
    inputencoding=utf8,
    frame=none,
    columns=[l]fullflexible,
    keepspaces=true,
    showstringspaces=false,
    mathescape=false,
}

\renewcommand{\paragraph}[1]{\noindent \textbf{#1}}

\DeclareUnicodeCharacter{2223}{$\mid$}
\DeclareUnicodeCharacter{211D}{$\mathbb{R}$}
\DeclareUnicodeCharacter{2081}{$_{1}$}
\DeclareUnicodeCharacter{2082}{$_{2}$}
\DeclareUnicodeCharacter{22A2}{$\vdash$}
\DeclareUnicodeCharacter{2115}{$\mathbb{N}$}
\DeclareUnicodeCharacter{2211}{$\sum$}
\DeclareUnicodeCharacter{2194}{$\leftrightarrow$}
\DeclareUnicodeCharacter{2208}{$\in$}
\DeclareUnicodeCharacter{1D55C}{$\mathbb{k}$}
\DeclareUnicodeCharacter{3B9}{$\iota$}
\DeclareUnicodeCharacter{3C0}{$\pi$}
\DeclareUnicodeCharacter{271D}{$\dagger$}
\DeclareUnicodeCharacter{2075}{$^{5}$}
\DeclareUnicodeCharacter{2074}{$^{4}$}
\DeclareUnicodeCharacter{2073}{$^{3}$}
\DeclareUnicodeCharacter{2072}{$^{2}$}
\DeclareUnicodeCharacter{2071}{$^{1}$}
\DeclareUnicodeCharacter{2200}{$\forall$}
\DeclareUnicodeCharacter{2A0D}{$\int$}
\usepackage[textsize=scriptsize]{todonotes}
\usepackage{cleveref}

\newcommand{\progresscount}{430\,}
\newcommand{\name}{\textsc{Lean}-\textsc{GAP}\xspace}

\newenvironment{red}{\relax\color{red}}{\hspace*{.5ex}\relax}
\newenvironment{blue}{\relax\color{blue}}{\hspace*{.5ex}\relax}
\newenvironment{lightgray}{\relax\color{lightgray}}{\hspace*{.5ex}\relax}
\newcommand{\ber}{\begin{red}}
\newcommand{\er}{\end{red}}
\newcommand{\beb}{\begin{blue}}
\newcommand{\eb}{\end{blue}}
\newcommand{\beg}{\begin{lightgray}}
\newcommand{\eg}{\end{lightgray}}

\definecolor{promptbg}{RGB}{248,249,250}
\definecolor{promptframe}{RGB}{210,214,220}
\definecolor{prompttitle}{RGB}{60,72,88}

\usepackage[most]{tcolorbox}
\usepackage{listings}
\usepackage{xcolor}

\definecolor{promptbg}{RGB}{248,249,250}
\definecolor{promptframe}{RGB}{210,214,220}

\newtcblisting{promptbox}[2][]{%
  breakable,
  enhanced,
  listing only,
  listing engine=listings,
  colback=promptbg,
  colframe=promptframe,
  coltitle=black,
  boxrule=0.5pt,
  arc=2mm,
  left=1.2mm,
  right=1.2mm,
  top=0.8mm,
  bottom=0.8mm,
  title={#2},
  fonttitle=\bfseries,
  attach boxed title to top left={xshift=2mm,yshift=-2mm},
  boxed title style={
    colback=white,
    colframe=promptframe,
    boxrule=0.5pt,
    arc=1mm,
  },
  listing options={
    language={},
    basicstyle=\ttfamily\small\color{black},
    keywordstyle=\color{black},
    commentstyle=\color{black},
    stringstyle=\color{black},
    identifierstyle=\color{black},
    emphstyle=\color{black},
    numberstyle=\color{black},
    showstringspaces=false,
    breaklines=true,
    breakindent=0pt,
    breakautoindent=false,
    columns=fullflexible,
    keepspaces=true,
    upquote=true,
    frame=none,
  },
  #1
}

\newtcolorbox{textbox}[1][]{
  breakable,
  enhanced,
  colback=promptbg,
  colframe=promptframe,
  boxrule=0.5pt,
  arc=2mm,
  left=1.2mm,
  right=1.2mm,
  top=0.8mm,
  bottom=0.8mm,
  before upper=\setlength{\parindent}{0pt},
  #1
}

\title{\name : A Dataset of Formalized Graduate Algebra Problems}
\author{%
{Seewoo Lee}\thanks{Equal contribution} \\ \texttt{\scriptsize seewoo5@berkeley.edu} 
\And
{Byung-Hak Hwang}\footnotemark[1] \\ \texttt{\scriptsize bhwang@kias.re.kr} 
\And
{Hyojae Lim} \\ \texttt{\scriptsize hlim@kias.re.kr} 
\And
{Jihoon Hyun} \\ \texttt{\scriptsize qawbecrdtey@kaist.ac.kr} 
\And
{Ilkyoo Choi} \\ \texttt{\scriptsize ilkyoo@hufs.ac.kr} 
\And
{Yeachan Park} \\ \texttt{\scriptsize ychpark@sejong.ac.kr} 
\And
{Jineon Baek} \\ \texttt{\scriptsize jineon@kias.re.kr} 
\And
{Hyukpyo Hong} \\ \texttt{\scriptsize hhong78@wisc.edu} 
\And
{Keewoo Lee} \\ \texttt{\scriptsize keewoo.lee@ethereum.org} 
\And
{Jaeseong Heo} \\ \texttt{\scriptsize hjs@hanyang.ac.kr} 
\And
{Hyungryul Baik} \\ \texttt{\scriptsize hrbaik@kaist.ac.kr} 
\And
{Chul-hee Lee}\thanks{Corresponding author} \\ \texttt{\scriptsize chlee@kias.re.kr} 
\And
{Kyu-Hwan Lee}  \\ \texttt{\scriptsize khlee@math.uconn.edu} } 

\begin{document}

\maketitle

\begin{abstract}
We present \name (Lean-\textbf{G}raduate \textbf{A}lgebra \textbf{P}roblems)\footnote{Dataset available at \href{https://www.kaggle.com/datasets/b8d166d6fecce97ae60db6e8a9560e6c015c7db50f609000f72d7cd05b70729d}{Kaggle}.},  \progresscount formalized graduate-level algebra problems from the textbook \emph{Abstract Algebra} by Dummit and Foote.
We develop a scalable pipeline consisting of PDF-to-\LaTeX{} preprocessing, autoformalization into Lean 4, and verification of informal--formal correspondence. While the preprocessing and autoformalization stages can be largely automated, we find that verification remains the most subtle and labor-intensive component, requiring careful human oversight.
Our contributions include (i) the construction of a structured dataset of formalized exercises, (ii) a systematic methodology for formalizing textbook mathematics, and (iii) an analysis of recurring challenges in the formalization process.
We also compare the performance of different autoformalization models and highlight key bottlenecks in translating informal statements into formal language.

\end{abstract}

\section{Introduction}
\newcommand{\SpecLogicRv}{\mathsf{\Psi}}
\newcommand{\TestSuiteRv}{\mathsf{T}}
\newcommand{\ProgramRv}{\mathsf{P}}
\newcommand{\ProgramSet}{\Pi}
\newcommand{\SpecNLRv}{\mathsf{S_{NL}}}
\newcommand{\SpecNLSet}{\mathcal{N}}
\newcommand{\SpecF}{\psi}
\newcommand{\SpecFFull}{\phi(x_1, \dots, x_n; \Program)}
\newcommand{\SpecNL}{\nu}
\newcommand{\SpecFSet}{\Psi}
\newcommand{\Program}{\pi}
\newcommand{\TestCase}{\tau}
\newcommand{\TestSet}{\mathcal{T}}

In recent years, the use of proof assistants, particularly \emph{Lean 4}, for formalizing mathematics has drawn considerable attention from both the mathematics and artificial intelligence (AI) research communities.
This interest has been driven in part by large-scale formalization efforts such as the Liquid Tensor Experiment \cite{scholze2019liquid}, the Prime Number Theorem \cite{strongpnt}, and Fermat's Last Theorem \cite{flt_formalization}, and by advances in automated reasoning and theorem proving exemplified by systems such as AlphaProof \cite{hubert2026olympiad}.
At the same time, the rapid development of large language models (LLMs) has brought the problem of \emph{autoformalization} to the forefront, with recent successes demonstrating the feasibility of translating informal mathematical arguments into fully formal proofs. In this context, large-scale formal libraries, most notably Lean's \texttt{Mathlib}~\cite{mathlib4}, play a central role, serving as extensive repositories of rigorously formalized mathematics and providing the foundational infrastructure upon which both human-driven and AI-assisted formalization efforts rely on.

Despite these advances, a substantial gap remains between informal mathematics and its formal counterparts. First, the volume of formalized mathematics remains small relative to the vast corpus of informal mathematical knowledge. Second, much of the recent progress has focused on competition-style or highly specialized benchmark problems, rather than standard, high-level material, particularly at the graduate level, that forms the foundation of mathematical training.
This latter issue presents a significant obstacle for the development of 
automated reasoning systems, as it reflects a scarcity of representative and structured datasets. An instructive analogy is that of attempting to train a student using only challenging problems, without first guiding them through the standard curriculum typically covered in the first years of graduate study. Such an approach neglects the systematic development of conceptual understanding and technical fluency.

The present work is motivated by the need to address this gap. Specifically, we initiate a systematic effort to formalize exercise problems from a widely used graduate-level textbook \emph{Abstract Algebra} by Dummit and Foote~\cite{dummit_foote_2003}. 
By doing so, we aim to construct a corpus of formalized mathematics that reflects standard mathematical practice, thereby providing both a foundation for training AI systems and a bridge between informal exposition and formal reasoning in research and education.
We carried out the project in several stages: (1) preprocessing a scanned PDF copy of the textbook into \LaTeX{} exercise statements, (2) translating the \LaTeX{} statements into Lean code via autoformalization, and (3) verifying the correctness of the resulting informal--formal pairs. Once the necessary infrastructure was in place, the first two steps required little human intervention. Moreover, the availability of multiple LLM-based autoformalization systems allowed us to compare the performance of different models within a unified pipeline.
Perhaps surprisingly to non-experts, the third step, i.e., verification, proved to be the most time-consuming, technical, and subtle. At present, there is no reliable automated method for checking whether a formal statement faithfully captures its informal counterpart. In particular, the mere fact that a piece of Lean code compiles successfully does not guarantee that it correctly represents the intended mathematical meaning of the original informal statement.
As a result, this stage required careful human oversight. The authors, mathematicians with PhDs and advanced graduate students, engaged in extensive cross-checking 
to identify errors, refine formulations, and resolve subtle ambiguities arising in the formalization process. It was striking to observe that even for standard graduate-level material, formalization demands complete understanding of the subject with a high degree of precision and sustained attention. This highlights both the current limitations of autoformalization systems and the importance of developing more robust methods for validating informal--formal correspondence.

Our contributions are threefold. First, we construct a dataset of formalized exercises drawn from a widely used textbook, naturally organized by topic, difficulty, and structural features;
we name the dataset \name{}.
Second, we develop a systematic pipeline for formalizing textbook exercises.
Third, we analyze recurring patterns and bottlenecks in the formalization process.
In the main body of the paper, we describe our methodology in detail. We compare the performance of different autoformalization models and analyze the types of difficulties and subtleties that arise in the faithful translation of informal statements into formal language. Representative examples are presented to illustrate these challenges, providing insights into both formalizing graduate-level mathematics and the development of AI-assisted theorem proving systems.

\section{Dummit--Foote exercises}
The textbook \emph{Abstract Algebra} by Dummit and Foote \cite{dummit_foote_2003} is one of the most widely adopted texts for graduate-level algebra courses. Designed to support a full-year course while containing substantially more material than is typically covered in that time, it offers a broad and carefully structured introduction to group theory, ring theory, module theory, field theory, Galois theory, homological algebra, commutative algebra, representation theory, and related advanced topics. Its breadth makes it both a foundational resource for students and a natural bridge to further study and research.

A distinguishing characteristic of the textbook is its extensive collection of exercises, which is an integral part of the exposition rather than merely supplementary problems. Many important ideas and results are introduced, developed, or foreshadowed through exercises, sometimes in carefully organized sequences that form a parallel narrative to the main text. From the perspective of formalization, this makes the exercises a natural and structured dataset for graduate-level algebra, but also presents challenges: their diversity in style and difficulty, their frequent reliance on implicit assumptions, and the informal introduction of concepts across sequences of problems all require careful interpretation when translating them into formal language.

Our goal is to formalize \emph{all} exercises from the textbook. 
In total, it contains 1,966 exercises. At the time of writing, we have formalized \progresscount problems, more than 20\% of these problems. As the project progresses, we plan to invite broader community participation to complete the remaining formalizations. The resulting dataset will be made publicly available;  permission from the copyright holder has already been granted.

\section{Formalization pipeline}
\subsection{PDF-to-\LaTeX{} preprocessing}
\label{sec:preprocessing}

The first stage of the pipeline is converting exercises from a scanned PDF
copy of the textbook into \LaTeX{} statements, since \LaTeX{} preserves mathematical
notation in a human-readable format.
We rendered the relevant pages as images using the \texttt{pdftoppm}
utility from the Poppler PDF rendering library\footnote{\href{https://poppler.freedesktop.org}{https://poppler.freedesktop.org}} and cropped out extraneous
elements in the bottom margin, including running section titles and page numbers.
The cropped images were then processed using the Mathpix Convert API\footnote{\href{https://mathpix.com/convert}{https://mathpix.com/convert}}, a third-party OCR service for STEM documents, to obtain initial \LaTeX{} transcriptions.

\subsection{\LaTeX{}-to-Lean autoformalization}
\label{sec:autoformalization}

The \LaTeX{} statements from the preprocessing stage are translated into formal specifications in Lean 4. We used a prompt-based approach with a large language model to convert structured mathematical language into syntactically correct Lean declarations while preserving their semantics. The prompt enforces consistent naming, avoids unnecessary auxiliary definitions, and ensures compatibility with \texttt{Mathlib}. The outputs are concise theorem statements with \lstinline{sorry} placeholders, serving as formal problem statements for subsequent verification and completion.

\begin{promptbox}{Prompt for autoformalization}
Give a Lean 4 code that formalizes the statement of the following algebra exercise.
Leave proof as `by sorry` and do not attempt to complete it.
Do not define extra variables, include them as statement of the formalized codes.
Assume `import Mathlib` is already in place and you do NOT need to import other submodules.
Keep your response short.
The theorem name has to be `DF_{sec}_{subsec}_{exercise_num}`.
If it consists of subproblems as (a), (b), (c), ..., then formalize each of them as `theorem DF_{sec}_{subsec}_{exercise_num}_a`, `theorem DF_{sec}_{subsec}_{exercise_num}_b`, etc.

Here's a statement of the exercise: {exercise_statement}

ONLY give Lean formalization WITHOUT any detailed explanations.
\end{promptbox}

\subsection{Verification of informal--formal correspondence}
\label{sec:verification}

LLM-based autoformalization is not guaranteed to be correct: a generated Lean file may elaborate successfully yet encode a statement that differs from the original exercise. Formalizing a statement without a proof is also
harder than it may look, because without a proof obligation we cannot
rely on the elaborator to expose a mismatch in meaning. We therefore
require every exercise in \name{} to pass a two-stage human review
before incorporating it into the dataset. Each exercise is checked by at
least two authors, a contributor who prepares it and a maintainer who
reviews it, so every single statement in the released
benchmark carries a mathematician's sign-off.

\begin{figure}[t]
\centering
\resizebox{\textwidth}{!}{%
\begin{tikzpicture}[
  x=22mm, y=14mm,
  font=\small,
  box/.style={draw, rounded corners=1.5pt, minimum height=10mm,
              minimum width=20mm, align=center, inner sep=2pt, fill=white},
  arr/.style={-{Stealth[length=1.8mm]}, thick},
  fback/.style={arr, dashed, draw=gray!60},
  lanehdr/.style={anchor=east, font=\small\itshape, align=right}
]

\begin{scope}[on background layer]
  \fill[blue!6]    (-0.55, 1.55) rectangle (5.55, 2.45);
  \fill[gray!10]   (-0.55, 0.55) rectangle (5.55, 1.45);
  \fill[orange!8]  (-0.55,-0.45) rectangle (5.55, 0.45);
\end{scope}

\node[lanehdr] at (-0.65, 2) {Human\\Contributor};
\node[lanehdr] at (-0.65, 1) {Automated\\(GitHub CI)};
\node[lanehdr] at (-0.65, 0) {Human\\Maintainer};

\node[box] (sel)  at (0, 2) {1.\ Select};
\node[box] (rew)  at (1, 2) {2.\ Informal\\rewrite};
\node[box] (form) at (2, 2) {3.\ Formalize\\(Lean~4)};
\node[box] (pr)   at (3, 2) {4.\ Open PR};

\node[box] (ci)   at (3, 1) {5a.\ Elaboration\\check};

\node[box] (rev)   at (4, 0) {5b.\ Semantic\\review};
\node[box] (merge) at (5, 0) {6.\ Merge};

\draw[arr] (sel)  -- (rew);
\draw[arr] (rew)  -- (form);
\draw[arr] (form) -- (pr);
\draw[arr] (pr)   -- (ci);
\draw[arr] (ci)   -- (rev);
\draw[arr] (rev)  -- (merge);

\draw[fback] (rev.north) -- (4, 2.75) --
    node[midway, above, font=\scriptsize]{revise}
    (2, 2.75) -- (form.north);
\end{tikzpicture}%
}
\caption{Per-exercise verification workflow. Each exercise is prepared
by a contributor (top lane), passes an automated elaboration check via
GitHub~CI (middle lane), and is reviewed for semantic correspondence by
an independent maintainer (bottom lane) before being merged. The
dashed arrow indicates a revision request by the maintainer that is returned to the contributor.}
\label{fig:workflow}
\end{figure}
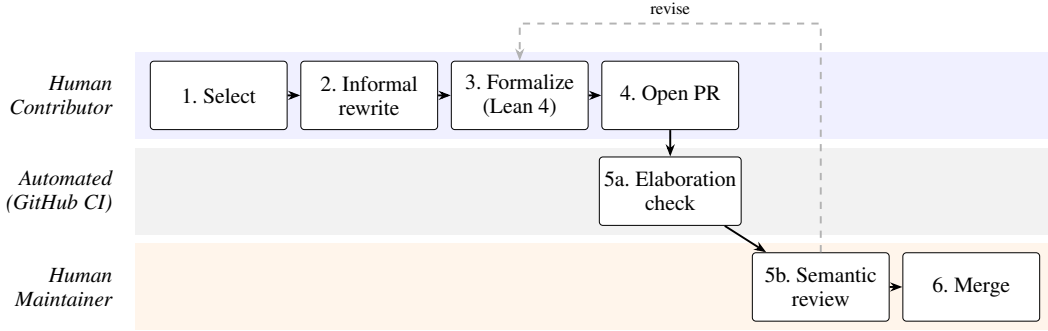

\paragraph{Per-exercise workflow.} Contributors work on exercises one
at a time through the following steps, illustrated in
Figure~\ref{fig:workflow}:
\begin{enumerate}
\item \emph{Selection.} The contributor picks an exercise that has not yet been formalized from the Git repository.
\item \emph{Informal rewrite.} The exercise is rephrased as a
  self-contained theorem statement rather than an instruction. For
  example, ``Show that $\mathrm{GL}_2(\mathbb{F}_2)$ is non-abelian.''
  is rewritten as ``$\mathrm{GL}_2(\mathbb{F}_2)$ is non-abelian.''
  Any notation or hypotheses inherited from the surrounding section is
  inlined so that the statement can be read without the textbook at
  hand.
\item \emph{Formalization.} The contributor produces a Lean~4 statement
  of the rewritten theorem, either by writing it directly or by
  starting from an LLM-generated candidate and editing it. In either
  case, the contributor is responsible for verifying that the final
  Lean statement is in exact correspondence with the informal rewrite.
\item \emph{Pull request.} The informal rewrite and the Lean statement
  are submitted together as a pull request to the project's 
  repository. Each pull request (PR) contains at most ten exercises, which will be examined by a
  maintainer.
\item \emph{Independent review.} Every PR is first checked
  automatically by a GitHub Actions workflow that elaborates each
  Lean declaration against the pinned \texttt{Mathlib} revision; a PR is only
  eligible for human review once this elaboration check passes. A maintainer then reads the informal--formal pair and
  focuses on semantic correspondence, verifying that the Lean
  declaration captures the mathematical content of the informal
  statement. Disagreements are resolved on the PR thread before
  merging.
\item \emph{Merge.} Once the maintainer approves, the PR is merged.
  The Lean declaration, together with the informal rewrite that was
  checked against it, becomes part of the released dataset.
\end{enumerate}

The repository is kept current with Lean and \texttt{Mathlib} updates: when a
new Lean version is released, or when \texttt{Mathlib} renames or refactors a
definition the dataset depends on, the affected declarations are
updated through the same PR-and-review process, so the human sign-off
is preserved across toolchain changes.

\section{Evaluation of autoformalization systems}
We evaluated autoformalization on \name from two angles: whether the generated Lean \emph{elaborates}, and whether the resulting statement \emph{matches} the informal exercise. Because no single model is reliable on both axes, we ran six systems in parallel: five single-shot baselines (GPT-5, Gemma4-31B (think), Goedel-Formalizer-V2-32B, Qwen3.6-35B-A3B, and DeepSeek-R1-Distill-Qwen-32B) and Codex, OpenAI's coding agent (GPT-5 backbone) wrapped in an autoformalization loop.\footnote{Snapshots used: \texttt{gpt-5-2025-08-07}, \lstinline{gemma4:31b-it-q4_K_M}, \lstinline{kdavis/goedel-formalizer-v2:32b}, \lstinline{qwen3.6:35b}, \lstinline{deepseek-r1:32b}.}
The single-shot systems use one API call per exercise with the prompt from Section~\ref{sec:autoformalization}. Codex instead retries each exercise with the failed code and Lean error message supplied as additional context, until elaboration succeeds or its iteration budget is exhausted, with no human intervention inside the loop. Because of compiler-grounded retries, Codex is not directly comparable to the single-shot baselines, but we include it as a reference point for what such loops can achieve on \name.

\subsection{Elaboration check}
\label{sec:elaboration}

Successful elaboration is necessary but not sufficient: a statement can elaborate yet be mathematically vacuous (e.g., conclusion \lstinline{True}, placeholder hypotheses \lstinline{(h : True)}, empty typeclasses, or an existential over \lstinline{True}) or fail to express the informal exercise. We complement it with eleven suspicious-statement patterns C1--C11 (Appendix~\ref{app:patterns}, per-model and per-pattern counts in Table~\ref{tab:susp-patterns}) and an LLM-based semantic check (Section~\ref{subsec:semantic-check}). Every generated \texttt{.lean} file was elaborated independently against the pinned Lean toolchain \lstinline{leanprover/lean4:v4.29.1} and \texttt{Mathlib} revision \lstinline{5e932f9}. We report outcomes per exercise, marking it \texttt{all\_pass} if every declaration elaborates, \texttt{any\_pass} if at least one does, \texttt{fail} if all attempts fail, and \texttt{missing} if the model produces no declaration. \lstinline{sorry} is allowed in proof position but counted as failure when used in a \lstinline{def} body or inside the statement itself.

\paragraph{Results.}
Table~\ref{tab:elab} reports per-exercise elaboration outcomes over all $1{,}966$ exercises.
GPT-5 leads the single-shot systems with $44.1\%$, followed by Goedel-Formalizer-V2-32B
at $36.2\%$, reflecting the benefit of autoformalization-specific fine-tuning.
The reasoning-trained Qwen3.6-35B-A3B and DeepSeek-R1-Distill-Qwen-32B underperformed
at $21.5\%$ and $5.4\%$; DeepSeek-R1 failed to emit any declaration for $1{,}040$
exercises, suggesting that chain-of-thought traces consume most of its output budget
before a clean Lean formalization is produced. Gemma4-31B (think) landed at $11.9\%$.
Codex reached $95.5\%$ any-pass, showing that compiler-grounded retries
close most of the gap left by single-shot generation. Failures are dominated
by \texttt{unknownIdentifier} (hallucinated names), \texttt{synthInstanceFailed}
(missing typeclasses), and \texttt{invalidField}; per-category counts are in Table~\ref{tab:err-cat}.

\begin{table}[t]
  \centering

  \caption{Per-exercise elaboration outcomes over all $1{,}966$ exercises (column definitions in text). Codex is listed as a reference rather than a directly comparable baseline.}
  \label{tab:elab}
  \vspace{.2cm}

  \begin{tabular}{lrrrr}
    \toprule
    Model & any\_pass & all\_pass & fail & missing \\
    \midrule
    GPT-5                          &  \textbf{867 (44.1\%)} &  577 &  942 &  157 \\
    Gemma4-31B (think)             &  233 (11.9\%) &  132 & \textbf{1733} &    0 \\
    Goedel-Formalizer-V2-32B       &  711 (36.2\%) &  \textbf{658} & 1253 &    2 \\
    Qwen3.6-35B-A3B                &  422 (21.5\%) &  306 & 1306 &  238 \\
    DeepSeek-R1-Distill-Qwen-32B   &  106 (5.4\%)  &   80 &  820 & \textbf{1040} \\
    \midrule
    Codex (agent loop)             & 1877 (95.5\%) & 1857 &   89 &    0 \\
    \bottomrule
  \end{tabular}
\end{table}

\begin{table}[t]
  \centering
  \caption{Elaboration error distribution by Lean category. Counts are failing declarations per category.}
  \label{tab:err-cat}
  \vspace{.2cm}
  
  \small
  \begin{tabular}{lrrrrr}
    \toprule
    Category & GPT-5 & Gemma-31B think & Goedel & Qwen3.6 & DeepSeek-R1 \\
    \midrule
    \texttt{lean.unknownIdentifier}      & 243 & 321 & 152 & 206 & 104 \\
    \texttt{lean.synthInstanceFailed}    & 237 & 218 & 334 & 212 & 138 \\
    \texttt{lean.invalidField}           &  53 &  90 &  82 &  81 &  38 \\
    \texttt{lean.dependsOnNoncomputable} &   1 &   3 &   0 &   4 &   0 \\
    other                                &   1 &   1 &   0 &   0 &   0 \\
    \bottomrule
  \end{tabular}
\end{table}

\paragraph{Cross-model behaviour and ceilings.}
Models disagreed on which exercises they could handle: among the $1{,}966$ exercises,
only $7$ were solved by every system and $654$ by exactly one,
a strong argument for running models in parallel rather than picking one.
The $60$ universally failed exercises cluster in advanced chapters ($39$ of $60$
in Chapters~10--18, with Ch.~12 and Ch.~17 alone contributing $13$ and $7$, respectively),
reflecting gaps in \texttt{Mathlib}'s coverage rather than model-specific weakness:
when the relevant definitions are missing, every model falls back on guessing
names Lean does not recognize. Even Codex's agent loop ceilings at $95.5\%$.
Many of the remaining $89$ failures stem not from missing mathematics but
from \texttt{Mathlib} drift since training (renamed lemmas,
restructured instances, definitions made noncomputable); it is a structural
limit for any pipeline whose LLM is fixed.

\subsection{Semantic check via LLM}
\label{subsec:semantic-check}

Elaboration and the C1--C11 patterns do not catch a statement that elaborates,
avoids every pattern, and still encodes the wrong mathematics: a quantifier
flipped, a wrong group named, a hypothesis silently weakened. We therefore
judged each generated \texttt{.lean} file with Claude Opus~4.7 against
a five-axis rubric: mathematical objects (S1), hypotheses (S2), conclusion (S3),
structure (S4: quantifiers and connectives), and specificity
(S5: concrete objects vs.\ arbitrary parameters).
The judge returns a faithfulness label (\texttt{faithful}/\texttt{partial}/\texttt{unfaithful}/\texttt{vacuous}),
an integer score from $0$ (unrelated) to $5$ (perfect), and zero or more issue codes
from a closed vocabulary. Every emitted file is judged regardless of elaboration,
so a faithful statement with a broken proof still receives a score.
The full prompt and code list are in Appendix~\ref{app:judge-prompt}.

\begin{table}[t]
  \centering
\caption{Per-file LLM-judge semantic evaluation. Mean score is $0$--$5$
($5$ = perfect, $3$ = main idea captured but some conditions missing, $0$ = vacuous or unrelated).
$N$ may exceed $1{,}966$ when a model splits an exercise across files,
or fall below when files are missing.}
  \label{tab:judge}
  \vspace{.2cm}
  \begin{tabular}{lrrrrrr}
    \toprule
    Model & N & Mean & Faithful & Partial & Unfaithful & Vacuous \\
    \midrule
    GPT-5                          & 1808 & \textbf{3.45} & \textbf{45.1\%} & 46.8\% &  5.9\% &  2.2\% \\
    Gemma4-31B (think)             & 1965 & 2.67 & 21.3\% & \textbf{61.6\%} & 15.8\% &  1.3\% \\
    Goedel-Formalizer-V2-32B       & 1962 & 2.59 & 24.3\% & 47.9\% & 22.5\% &  5.3\% \\
    Qwen3.6-35B-A3B                & 1727 & 2.76 & 25.8\% & 54.7\% & 17.3\% &  2.3\% \\
    DeepSeek-R1-Distill-Qwen-32B   &  926 & 1.15 &  2.7\% & 28.8\% & \textbf{48.3\%} & 20.2\% \\
    \midrule
    Codex (agent loop)             & 1989 & 3.56 & 51.2\% & 38.4\% &  5.8\% &  4.6\% \\
    \bottomrule
  \end{tabular}
\end{table}

\paragraph{Results.}
Per-file outcomes are in Table~\ref{tab:judge}. GPT-5 leads the single-shot models, consistent with its elaboration ranking. However, its mean score is still far from perfect, indicating that many outputs only partially capture the target statement or omit important conditions. The results also show that syntactic success is not equivalent to semantic correctness: Goedel-Formalizer-V2-32B has the highest all-pass count among single-shot models, but a lower semantic mean and higher unfaithful rate than GPT-5 and Qwen3.6-35B-A3B.

\paragraph{Limitations.}
The judge is itself an LLM: its absolute scores are not calibrated against expert review, and it can mis-rate edge cases. We therefore read Table~\ref{tab:judge} as a relative signal across models, not as a verdict on individual exercises. Human review by a contributor and a maintainer (Section~\ref{sec:verification}) remains the reference for the released benchmark.

\section{Challenges in verifying informal--formal correspondence}
The maintainer review described in this section is
where verification was performed,
and it is the step we found most labor-intensive.
Two kinds of difficulty recurred during this stage.
The first is intrinsic to the exercise:
some problems rely on mathematical content that resists clean translation into Lean.
The second is detection:
a Lean file may elaborate, avoid every suspicious pattern of Section~\ref{sec:elaboration},
and still encode the wrong mathematics;
we found recent AI assistants useful as second readers in this case.

\subsection{Examples of exercises that are hard to formalize}

\subsubsection{Exercises with geometric arguments}

There are several exercises that include geometry, which are hard to formalize in Lean in general.
For example, Exercises 13--18 of Chapter 14.5 are on a construction of a regular 17-gon.
The first two exercises (13 and 14) are purely algebraic problems, while the next exercises are on plane geometry and the actual construction of a regular 17-gon, including several figures of intermediate steps.
It is not easy to formalize the notion of a straightedge and compass construction in Lean, which would require a significant amount of effort to do it properly\footnote{There is a partial formalization of the notion of constructible numbers by Barroero, Bishop, Brasca, and Zilberberg, which can be found at \href{https://github.com/riccardobrasca/constructible}{github.com/riccardobrasca/constructible}. They defined \lstinline{IsConstructible K L} for given fields $K \subseteq L$ as an inductive type, but did not show the correspondence between this notion of constructibility and geometric moves by a straightedge and compass.}.

\subsubsection{Exercises requiring answer-embedded formalization}
Some exercises require the solver to determine all possible answers or to enumerate all (counter)examples satisfying a given condition. Such problems pose a subtle challenge in formalization, as the statement of the exercise may need to incorporate aspects of the solution itself. For instance, Exercise 11 of Chapter 1.1 asks for the order of each element in $\mathbb{Z}/12\mathbb{Z}$, and GPT-5 formalized it as follows. Here we show only the first three elements:
\begin{lstlisting}[language=Lean, style=leanplain]
theorem DF_1_1_11 :
    addOrderOf (0 : ZMod 12) = 1 ∧
    addOrderOf (1 : ZMod 12) = 12 ∧
    addOrderOf (2 : ZMod 12) = 6 ∧
    ...
\end{lstlisting}
This formalized statement already incorporates the answers for each element and asks the solver only to prove their correctness. This approach is also adopted by several existing benchmarks, including MiniF2F \cite{zheng2021minif2f}, FIMO \cite{liu2023fimo}, Lean-Workbook \cite{ying2024lean}, and DeepSeek-ProverBench \cite{ren2025deepseekprover-v2}. However, for benchmarking purposes, this formulation does not assess a model's ability to ``discover'' answers; rather, it evaluates only whether the model can verify that a given answer is correct.

Instead, we decided to leave the answers as definitions with \lstinline{sorry}s to be filled in, as follows:
\begin{lstlisting}[language=Lean, style=leanplain]
def DF_1_1_11_ans : Fin 12 → ℕ
    | 0 => sorry
    | 1 => sorry
    | 2 => sorry
    ...

theorem DF_1_1_11 : ∀ i : ZMod 12, addOrderOf i = answer i := by
\end{lstlisting}
This is better than the previous formalization in the sense that it asks the model to both fill in the answers and prove that they are correct, and some of the previous benchmarks including PutnamBench \cite{tsoukalas2024putnambenchevaluatingneuraltheoremprovers} follow this convention. However, one can easily hack the task by defining \lstinline{answer i} as \lstinline{addOrderOf (i : ZMod 12)} instead of providing the expected numerical answers. Then the exercise can be proved simply by \lstinline{intro i; rfl}, which is far from what we intended to test.
To mitigate this issue, we wrote a Python script that checks whether a submitted answer satisfies problem-specific specifications. If the check passes, then the script replaces the \lstinline{sorry}s with the submitted answers and verifies that the resulting proof is valid. 
Note that some previous datasets use different approaches, such as \texttt{SymPy}-based checks \cite{lewkowycz2022solving,glazer2024frontiermath} or formal verification with restricted tactics \cite{liu2025beyond}.

\subsubsection{Exercises requiring definitions not yet in \texttt{Mathlib}}
Some of the exercises require definitions that are not yet formalized in \texttt{Mathlib}.
Exercise 6 of Chapter 14.8 asks to prove that the Galois group of $x^5+15x+12$ (over $\mathbb{Q}$) is $F_{20}$, the Frobenius group of order $20$, and $F_{20}$ is not formalized yet in \texttt{Mathlib}.
So we define it via generators and relations following the definition in Chapter 5.3 and state the exercise against
this construction:
\begin{lstlisting}[language=Lean, style=leanplain]
inductive Generators | u | v deriving DecidableEq, Repr

open FreeGroup in
def F₂₀ := PresentedGroup {
  (of Generators.u)^4,
  (of Generators.v)^5,
  (of Generators.u) * (of Generators.v) * (of Generators.u)^(-1 : ℤ) * (of Generators.v)^(-2 : ℤ)
}

deriving instance Group for F₂₀

theorem DF_14_8_6 : Nonempty ((X ^ 5 + 15 * X + 12 : ℚ[X]).Gal ≃* F₂₀) := by
\end{lstlisting}

There are several other exercises that are hard to formalize; these can be found in Appendix \ref{subsec:appendix_additional_examples}.

\subsection{Finding and fixing misformalized exercises with AI}

We also used Claude Opus 4.7 and Aristotle \cite{achim2025aristotle} to help with correcting autoformalized statements, or even human-formalized statements that contained errors.
We first illustrate an example with a semantic error. Exercise 15 of Chapter 1.1 asks to prove $(a_1 a_2 \cdots a_n)^{-1} = a_n^{-1} a_{n-1}^{-1} \cdots a_1^{-1}$ for elements $a_1, a_2, \dots, a_n$ of a group $G$.
The initial hand-crafted formalization was the following:
\begin{lstlisting}[language=Lean, style=leanplain]
theorem DF_1_1_15 {G : Type*} [Group G] (l : List G) :
    (l.foldl (· * ·) 1)⁻¹ = (l.map (·⁻¹)).foldr (· * ·) 1 := by    
\end{lstlisting}
Claude Opus 4.7 found that the right-hand side gives $a_1^{-1} a_2^{-1} \cdots a_n^{-1}$ instead, and also suggested correcting it to
\begin{lstlisting}[language=Lean, style=leanplain]
theorem DF_1_1_15 {G : Type*} [Group G] (l : List G) :
    (l.foldl (· * ·) 1)⁻¹ = (l.reverse.map (·⁻¹)).foldl (· * ·) 1 := by    
\end{lstlisting}

Notably, we found a subtle case where an autoformalized statement had neither a syntax nor a semantic error, but an original informal statement was faulty. Exercise 21 of Chapter 2.3 is the following:
\begin{textbox}
    Let $p$ be an odd prime and \textbf{let $n$ be a positive integer}. Use the binomial theorem to show that $(1 + p)^{p^{n-1}} \equiv 1 \pmod{p^n}$ but $(1+p)^{p^{n-2}} \not\equiv 1 \pmod{p^n}$. Deduce that $1 + p$ is an element of order $p^{n-1}$ in the multiplicative group $(\mathbb{Z} / p^{n}\mathbb{Z})^\times$.
\end{textbox}
The original statement above requires an additional constraint: $n\ge 2$ is necessary in order for 
$p^{n-2}$ to be well-defined as an integer. Also, this additional constraint is crucial to make the statement true in Lean.
The following initial formalization closely followed the original statement:
\begin{lstlisting}[language=Lean, style=leanplain]
theorem DF_2_3_21_1 {p n : ℕ} (hp : p.Prime) (hodd : Odd p) (hn : 0 < n) :
    (1 + p) ^ (p ^ (n - 1)) ≡ 1 [ZMOD (p ^ n)] ∧
    ¬ (1 + p) ^ (p ^ (n - 2)) ≡ 1 [ZMOD (p ^ n)] := by
\end{lstlisting}
Aristotle \cite{achim2025aristotle} was able to find a counterexample, namely $(n, p) = (1, 3)$, showing that the statement above is false. In Lean, for natural numbers $a, b$ with $a < b$, the expression $a - b$ is defined to be $0$. Hence $1 - 2 = 0$, and the pair $(n, p) = (1, 3)$ indeed gives a counterexample, which is not aligned with the original intent of the problem. It is now corrected with \lstinline{hn : 2 ≤ n}.

\section{Related work}
\subsection{Autoformalization} 

Autoformalization is the task of automatically translating informal mathematical statements into formal representations in interactive theorem provers such as Lean. Recent advances in large language models have led to notable progress, with approaches broadly falling into sequence-to-sequence translation, retrieval-augmented generation, and compiler- or process-guided methods \cite{azerbayev2023proofnet, liu2025rethinking, Lu2024PDA}. In addition, large-scale synthetic data generation has been shown to improve performance and scalability \cite{xin2024deepseekprover}. Domain-specific studies further demonstrate that structural constraints can make autoformalization more tractable, while also highlighting the importance of domain knowledge and dataset design \cite{Kaiyu2024AEG}.

Verification of informal--formal correspondence (i.e., semantic verification) remains a fundamental challenge in autoformalization, due to the ambiguity and implicit assumptions in informal mathematics and the lack of reliable evaluation metrics. FormalMATH~\cite{yu2025formalmath} reports that 60.7\% of syntactically valid Lean 4 statements were rejected at the semantic verification stage, revealing a substantial gap between syntactic correctness and semantic fidelity. Prior works have proposed proxy metrics including type-check rate~\cite{azerbayev2023proofnet}, back-translation-based semantic consistency and symbolic equivalence~\cite{li2024autoformalize}, cross-provability-based metrics~\cite{liu2025rethinking}, and operator-tree structural similarity~\cite{liu2026assess}. Poiroux et al.~\cite{poiroux2025reliable} provided a more systematic treatment via BEq+ and accompanying benchmarks. Despite these efforts, reliable semantic evaluation remains an open problem~\cite{liu2025rethinking,poiroux2025reliable}. Furthermore, scaling autoformalization beyond isolated statements to textbook- or project-level settings further compounds these challenges through dependency management and global consistency requirements~\cite{Amaury2026ATF,Wen2026M2F}.

\subsection{Datasets and benchmarks for formal mathematics}

Datasets for formal mathematics span a range from high-school Olympiad benchmarks such as miniF2F \cite{zheng2021minif2f} and FIMO \cite{liu2023fimo} to (under)graduate and research-oriented challenges including ProofNet~\cite{azerbayev2023proofnet}, TaoBench \cite{taylor2026taobench}, PutnamBench \cite{tsoukalas2024putnambenchevaluatingneuraltheoremprovers}, FormalQualBench \cite{FormalQualBench}, and the FATE series \cite{jiang2026fate,shen2025realprover}. Notably, both ProofNet \cite{azerbayev2023proofnet} and FATE-M \cite{shen2025realprover} specifically target Dummit--Foote \cite{dummit_foote_2003} as a primary source to represent the core (under)graduate algebra curriculum. ProofNet formalizes 87 problems from this textbook through expert-led manual formalization in Lean 3. In contrast, FATE-M consists of 141 problems in total, curated from 12 leading textbooks including Dummit--Foote, and employs student-led formalization in Lean 4 followed by rigorous PhD-level review to provide compiler-verified formal proofs. Recent benchmarks have expanded in both mathematical scope and system support.
For instance, PutnamBench provides a multi-language collegiate competition benchmark supporting Lean 4, Isabelle, and Coq, while FATE-X \cite{jiang2026fate} addresses advanced commutative algebra at PhD qualifying exam levels that can exceed current library coverage.

To address data scarcity, construction methodologies have shifted from manual formalization to large-scale automated pipelines. DeepSeek-Prover \cite{xin2024deepseekprover} and Lean Workbook \cite{ying2024lean} utilize synthetic generation and active learning to generate large-scale corpora of formal statements and proofs. Complementary efforts such as HERALD \cite{gao2025herald} and FORML4 \cite{Lu2024PDA} augment formal libraries with natural language annotations and leverage compiler feedback for process-driven autoformalization. These approaches aim to improve semantic alignment between informal reasoning and formal proof steps.

\section{Conclusion}
In this work, we have initiated a systematic effort to bridge the gap between informal mathematical practice and its formal counterparts by focusing on standard graduate-level material in abstract algebra. Through the formalization of exercises from \emph{Abstract Algebra} by Dummit and Foote \cite{dummit_foote_2003}, we have constructed a dataset of graduate algebra problems, called \name, that reflects the structure, breadth, and level of difficulty typical of mainstream mathematical training. 
While recent advances in autoformalization and LLMs have enabled substantial automation in the early stages of the pipeline, current LLMs cannot be relied upon as end-to-end autoformalizers, even for standard graduate-level material: a model can elaborate against \texttt{Mathlib} at a high rate yet still omit hypotheses or capture only part of the informal claim. Verification of informal--formal correspondence therefore remains a fundamental bottleneck, requiring careful human oversight and adequate mathematical understanding throughout the pipeline.

Rather than treating autoformalization as a problem to be solved end-to-end by a single LLM, our pipeline is built around productive human--AI collaboration. AI assistants serve as useful second readers during verification, identifying subtle errors in hand-crafted formalizations and surfacing unstated hypothesis in the textbook itself. Our pipeline thus offers an efficient template for deploying AI as collaborators within a human-in-the-loop process, rather than as standalone formalizers.

Beyond the construction of a dataset, our work develops a systematic methodology for formalizing textbook mathematics and provides a detailed analysis of the challenges that arise in this process. The insights gained from this study, ranging from ambiguities in informal statements to structural mismatches between human reasoning and formal systems, suggest that progress depends on the joint evolution of the original mathematics material, the underlying formal library, the evaluation infrastructure, and the model itself, with implications for both formalization of standard graduate-level mathematics and the development of AI-assisted theorem proving tools.

Looking ahead, we plan to complete the formalization of the remaining exercises through broader community participation, creating a comprehensive benchmark for graduate-level mathematics.  We also plan to use \name to generate informal--formal pairs of solutions via automated theorem provers, offering a complementary evaluation framework to existing benchmarks focused on competition-style problems. By focusing on standard mathematical training, this direction may be viewed as a step toward developing an ``AI mathematics student,'' if not yet an ``AI mathematics researcher.''  \name will also serve as an educational resource: making formalized versions of standard exercises widely available can support learning, curriculum development, the integration of formal methods into mainstream mathematical practice, and deeper interactions between the mathematics and AI communities.

\begin{ack}
We are grateful to John Wiley \& Sons, Inc. for granting us permission to use material from \emph{Abstract Algebra} by Dummit and Foote \cite{dummit_foote_2003} in this work, in the form of transformed representations as computer code, for non-commercial research and educational purposes.
We thank Wyatt Thompson from OpenAI, who generously provided credits for GPT-5 to the first-named author. We also thank Nayeong Kim for contributions to formalizing problems in the dataset.
Ilkyoo Choi was supported by the Hankuk University of Foreign Studies Research Fund, the National Research Foundation of Korea (NRF) grant funded by the Korea government (MSIT) (RS-2025-23324220), Institute for Basic Science (IBS-R029-C1), and the Korea Institute for Advanced Study (KIAS) grant funded by the Korea government.
Hyungryul Baik was supported by the National Research Foundation of Korea (NRF) grant funded by the Korean government (MSIT)
(No.\ RS-2025-00513595). 
Byung-Hak Hwang and Hyojae Lim were supported by KIAS Individual Grant (AP098201, AP109701) via the Center for Artificial Intelligence and Natural Sciences at Korea Institute for Advanced Study (KIAS). 
This work was supported by the Center for Advanced Computation (CAC) at Korea Institute for Advanced Study (KIAS).
\end{ack}

\bibliography{references}
\bibliographystyle{references}

\appendix
\section{Appendix}
\subsection{Suspicious-statement patterns}
\label{app:patterns}

Successful Lean elaboration is a necessary but not a sufficient condition
for a correct autoformalization: a file can type-check and still encode a
mathematically vacuous statement. We therefore pair the elaboration check
of Section~\ref{sec:elaboration} with the eleven suspicious-statement
patterns (C1--C11) defined below. Each pattern is detected with simple
checks on the generated \texttt{.lean} files: regular expressions on the
source together with a few queries on the parsed declaration. A
declaration that elaborates but matches at least one pattern is deducted
from the \emph{strict} pass rate. Patterns C1, C2, C4--C8 pass
elaboration silently because Lean has no reason to reject a trivially-true
statement, so unless we check for them they would be counted as successes.
Patterns C3 and C9--C11 usually make elaboration fail, but a few files
still slip past it (for example, by hiding \lstinline{sorry} inside a
\lstinline{def} whose declared type happens to match), so we list them as
well. The pattern list was calibrated against the human-curated reference
set (Section~\ref{sec:elaboration}) and the full classification is
released with the dataset.

\begin{description}[leftmargin=2.4em,labelindent=0pt,itemsep=2pt,topsep=4pt,parsep=0pt,font=\normalfont\bfseries]
  \item[C1.] Trivial conclusion \lstinline{: True}: the theorem claims \lstinline{True} and preserves none of the informal content.
  \item[C2.] Answer-as-free-parameter: a parameter named \lstinline{answer}, \lstinline{cond}, \lstinline{result}, etc.\ appears on both sides of a biconditional (\lstinline{(answer : Prop) : M ∈ B ↔ answer}); vacuously true under \lstinline{answer := M ∈ B}.
  \item[C3.] Placeholder text: tokens such as \lstinline{some_condition}, \lstinline{/* conditions here */}, \lstinline{_to_be_filled}, or \lstinline{Please replace} remain in the code.
  \item[C4.] Placeholder hypothesis \lstinline{(h : True)}: a condition from the informal statement is encoded as a \lstinline{True} hypothesis, weakening ``if \(X\) then \(Y\)'' into an unconditional claim.
  \item[C5.] \lstinline{by trivial} on \lstinline{: True}: a C1-style conclusion is paired with a completed trivial proof, so the declaration carries no \lstinline{sorry} warning.
  \item[C6.] \lstinline{def ... : Prop := True}: the central concept is defined as \lstinline{True}; every theorem mentioning it is vacuously satisfied, with the failure hidden one layer below the theorem.
  \item[C7.] Vacuous existential: a statement of the form \lstinline{∃ x, True} (or \lstinline{∃ x, x = x}, or an existential over an unconstrained \lstinline{Prop} binder) with no additional constraint on the witness.
  \item[C8.] Excluded middle as conclusion: the conclusion is a disjunction \lstinline{P ∨ ¬P}, always provable in classical logic regardless of \(P\).
  \item[C9.] \lstinline{def X := sorry}: a helper definition is skipped entirely; dependent theorems elaborate but have no mathematical content.
  \item[C10.] \lstinline{sorry} inside a statement's \lstinline{let}: a \lstinline{let}-binding within the theorem statement has \lstinline{:= sorry} as the defined value.
  \item[C11.] \lstinline{sorry} as right-hand side of $\leftrightarrow$: the statement includes \lstinline!... ↔ sorry!; one side of the biconditional is literally absent.
\end{description}

\paragraph{Per-model distribution.}
Profiles differ across models (Table~\ref{tab:susp-patterns}): DeepSeek-R1
falls back on trivial \lstinline{: True} conclusions (C1) and proofs (C5),
Goedel and GPT-5 concentrate on vacuous existentials (C7), and
Gemma4-31B's few flags are mostly \lstinline{def ... := sorry} (C9).

\begin{table}[t]
  \centering
  \caption{Suspicious-pattern statistics by single-shot model. Upper block:
  per-pattern occurrences (a single declaration matching two patterns
  contributes to both rows). Lower block: total declarations submitted by
  the model and the share matching at least one of C1--C11. Codex is
  omitted because its agent loop changes the baseline at which
  pattern-level comparisons are meaningful.}
  \label{tab:susp-patterns}
  \vspace{.2cm}
  \small
  \setlength{\tabcolsep}{4.5pt}
  \begin{tabular}{llrrrrr}
    \toprule
    ID & Pattern & GPT-5 & Gemma-31B & Goedel & Qwen3.6 & DeepSeek-R1 \\
    \midrule
    C1     & trivial conclusion \lstinline!: True!         & 12 &  3 &  0 &  5 & 85 \\
    C2     & answer-as-free-parameter                      &  0 &  0 &  0 &  1 &  0 \\
    C3     & placeholder text                              &  0 &  1 &  0 &  0 &  5 \\
    C4     & placeholder hypothesis \lstinline!(h : True)! &  7 &  1 &  1 &  1 &  0 \\
    C5     & \lstinline!by trivial! on \lstinline!: True!  &  5 &  0 &  0 &  0 & 37 \\
    C6     & \lstinline!def ... : Prop := True!            & 14 &  0 &  0 &  0 &  0 \\
    C7     & vacuous existential                           & 54 &  6 & 66 & 27 &  2 \\
    C8     & excluded middle as conclusion                 &  1 &  0 &  0 &  0 &  0 \\
    C9     & \lstinline!def X := sorry!                    &  0 & 12 &  8 &  0 &  0 \\
    C10/11 & \lstinline!sorry! inside statement            &  0 &  1 &  1 &  3 &  0 \\
    \midrule
    \multicolumn{2}{l}{Total occurrences}                 &   93 &   24 &   76 &   37 &  129 \\
    \midrule
    \multicolumn{2}{l}{Total declarations}                & 4679 & 3858 & 2334 & 3340 & 2009 \\
    \multicolumn{2}{l}{Flag rate}                         & 1.9\% & 0.6\% & 3.3\% & 1.1\% & 4.6\% \\
    \bottomrule
  \end{tabular}
\end{table}


\subsection{The LLM judge prompt}
\label{app:judge-prompt}

The verbatim prompt used by the judge of
Section~\ref{subsec:semantic-check}. Each (exercise, model) pair is
sent this text as the system message, followed by the informal
\LaTeX{} statement and the generated Lean file.

\begin{promptbox}{Prompt for LLM-judge semantic check}
You are an expert mathematician evaluating Lean 4 formalizations of exercises from Dummit & Foote's "Abstract Algebra". Your task is to judge whether each Lean statement faithfully represents the original exercise.

IMPORTANT RULES:
- Focus on the STATEMENT only. `sorry` in proof bodies is expected -- ignore it.
- `sorry` in definition bodies (def X := sorry) or let bindings IS a problem -- it means the definition wasn't provided.
- `: True` as a conclusion means the formalization gave up -- it's vacuous (score 0).
- `def X : Prop := True` is a vacuous definition that makes any theorem using X meaningless.
- Helper definitions (def, class, instance) before the main theorem are part of the formalization. Evaluate them together.
- Sub-part theorems (DF_X_Y_Z_a, _b, _c, _1, _2) together cover one exercise -- evaluate them as a group.
- Docstrings (/-- ... -/) are just comments, ignore them for evaluation.

For each exercise, evaluate on these 5 criteria:

S1. Mathematical Objects: Are the structures (groups, rings, matrices, etc.) correctly represented?
S2. Hypotheses: Are all conditions from the original captured? Any missing or extra?
S3. Conclusion: Does the Lean conclusion match what the exercise asks to prove/determine?
S4. Structure: Are quantifiers (forall / exists), connectives (-> / <-> / and / or) used correctly?
S5. Specificity: If the exercise names concrete objects (S_3, Z/nZ, etc.), are they concrete in the formalization (not arbitrary parameters)?

Return a JSON array of objects, one per exercise. Each object:
{
  "id": "<exercise_id__model>",
  "faithfulness": "faithful" | "partial" | "unfaithful" | "vacuous",
  "score": <0-5>,
  "issues": [<list of issue codes>],
  "reasoning": "<1-3 sentences>"
}

Score guide:
  5 = perfect formalization (possibly with minor style differences)
  4 = minor issue (equivalent but unusual formulation)
  3 = main idea captured but missing some conditions
  2 = partially correct, significant issues
  1 = barely related to original
  0 = vacuous / completely wrong / unrelated

Issue codes (closed vocabulary):
  missing_hypothesis       -- a condition from the original is absent
  extra_hypothesis         -- hypothesis not in the original is added
  wrong_conclusion         -- the conclusion doesn't match the exercise
  free_variable_distortion -- concrete objects replaced by arbitrary parameters
  vacuous_statement        -- conclusion is trivially true (e.g., True)
  wrong_structure          -- structural encoding is inappropriate
  wrong_quantifier         -- forall / exists direction is wrong
  missing_typeclass        -- required typeclass (e.g., [Group G]) is missing
  conclusion_too_weak      -- conclusion is weaker than what the exercise asks
  conclusion_too_strong    -- conclusion is stronger than what the exercise asks
  wrong_object             -- wrong mathematical object used
  answer_as_parameter      -- answer is left as a free parameter instead of being determined
  def_body_sorry           -- definition body uses sorry (not proof body)
  wrong_hypothesis         -- a hypothesis is stated incorrectly

Return ONLY the JSON array, no other text.
\end{promptbox}



\subsection{Per-model issue distribution}
\label{app:judge-issues}

Table~\ref{tab:judge-issues} reports the per-model count of each
issue code emitted by the judge of
Section~\ref{subsec:semantic-check} across all judged files. Rows
are sorted by total occurrences. A single file can carry multiple
issue codes, so column counts can exceed the per-model file counts
$N$ in Table~\ref{tab:judge}.

The profiles differ noticeably across models: for example,
Gemma-31B leans heavily on \texttt{wrong\_structure} and \texttt{wrong\_object} while
Codex concentrates on \texttt{conclusion\_too\_weak} and
\texttt{missing\_hypothesis}, reflecting the distinct ways each
model fails.

\begin{table}[t]
  \centering
  \caption{Issue-code occurrences per model. A judged file matching
  multiple codes contributes to each row independently.}
  \label{tab:judge-issues}
  \vspace{.2cm}
  \small
  \begin{tabular}{lrrrrrr}
    \toprule
    Issue code & GPT-5 & Gemma-31B & Goedel & Qwen3.6 & DeepSeek-R1 & Codex \\
    \midrule
    \texttt{wrong\_structure}            & \textbf{331} & \textbf{899} & \textbf{649} & \textbf{536} & \textbf{409} & 215 \\
    \texttt{wrong\_object}               & \textbf{253} & \textbf{725} & \textbf{564} & \textbf{696} & \textbf{378} & 120 \\
    \texttt{wrong\_conclusion}           & \textbf{235} & 370 & \textbf{505} & 390 & \textbf{500} & 176 \\
    \texttt{missing\_hypothesis}         & 174 & 223 & 304 & 242 & \textbf{391} & \textbf{305} \\
    \texttt{conclusion\_too\_weak}       & 189 & 103 & 207 & 102 &  45 & \textbf{338} \\
    \texttt{vacuous\_statement}          & 137 &  45 & 133 &  53 & 205 & 110 \\
    \texttt{free\_variable\_distortion}  &  95 &  64 &  91 &  93 & 174 & 154 \\
    \texttt{wrong\_hypothesis}           &  90 & 120 & 167 & 129 &  53 &  77 \\
    \texttt{missing\_typeclass}          &  31 & 143 &  74 &  73 & 107 &  28 \\
    \texttt{extra\_hypothesis}           &  52 &  32 &  41 &  68 &  44 &  49 \\
    \texttt{answer\_as\_parameter}       &  28 &  38 &  96 &  51 &  19 &  45 \\
    \texttt{def\_body\_sorry}            &  34 &  43 &  60 &  42 &  13 &  21 \\
    \texttt{wrong\_quantifier}           &   5 &  17 &  15 &   6 &  11 &   5 \\
    \texttt{conclusion\_too\_strong}     &   7 &   2 &   9 &   7 &   0 &   4 \\
    \bottomrule
  \end{tabular}
\end{table}


\subsection{Additional exercise examples}
\label{subsec:appendix_additional_examples}

\paragraph{Exercises depending on other statements}

Several exercises depend on, or explicitly refer to, other exercises. For example, Exercise 16 of Chapter 3.2 asks the reader to prove Fermat's little theorem, i.e. $a^p \equiv a \pmod{p}$ for all prime $p$ and integer $a$, \emph{using Lagrange's theorem}. If one ignores the reference to Lagrange's theorem, the exercise can be formalized as follows:
\begin{lstlisting}[language=Lean, style=leanplain]
theorem DF_3_2_16 {p : ℕ} (hp : p.Prime) (a : ℤ) : a ^ p ≡ a [ZMOD p] := by
  sorry
\end{lstlisting}
However, another possibility is to formalize Lagrange's theorem separately and include it as an assumption in the statement of the exercise:
\begin{lstlisting}[language=Lean, style=leanplain]
def DF_3_2_16_Lagrange : Prop := ∀ (G : Type*) [Group G] [Finite G], ∀ (H : Subgroup G),
    Nat.card H ∣ Nat.card G

theorem DF_3_2_16' (h : DF_3_2_16_Lagrange) {p : ℕ} (hp : p.Prime) (a : ℤ) :
    a ^ p ≡ a [ZMOD p] := by
  sorry
\end{lstlisting}
The intended solution is to apply Lagrange's theorem to the multiplicative group $(\mathbb{Z} / p\mathbb{Z})^\times$ and the subgroup generated by $a$ (and consider the $a \equiv 0 \pmod{p}$ case separately), and an AI system may follow this approach when proving \lstinline{DF_3_2_16'}. However, there are also multiple proofs of Fermat's little theorem that do not use Lagrange's theorem, such as proofs using the binomial theorem or a combinatorial argument. Thus, it is possible to prove \lstinline{DF_3_2_16} without using such a hypothesis; in fact, this result already exists in \texttt{Mathlib} as \lstinline{Nat.ModEq.pow_card_sub_one_eq_one}. We therefore include both versions of the formalization, one with the Lagrange theorem hypothesis and one without it.

Another example is Exercise 5 of Chapter 4.4, which asks the reader to prove that the automorphism group of the dihedral group $D_4$ of order $8$ is isomorphic to $D_4$, \emph{using the fact that $D_4 \trianglelefteq D_8$}.\footnote{Dummit--Foote denote the order-$8$ dihedral group by $D_8$. Here we follow the convention in which $D_n$ denotes the dihedral group of order $2n$, rather than $D_{2n}$; this also agrees with the convention used in \texttt{Mathlib}.}
The intended solution is as follows: if $D_4$ is known to be a normal subgroup of $D_8$, then $D_8$ acts on $D_4$ by conjugation, giving a homomorphism $D_8 \to \mathrm{Aut}(D_4)$. Computing its kernel, which is the centralizer of $D_4$ in $D_8$, and its image then gives the desired isomorphism.

However, it is unclear whether the mere existence of an embedding $\phi : D_4 \to D_8$ with normal image is sufficient for the intended formalization, or whether one must give an explicit description of $\phi$ in terms of generators. Therefore, we decided to formalize only the version without the hint, as follows:
\begin{lstlisting}[language=Lean, style=leanplain]
theorem DF_4_4_5 : Nonempty ((MulAut (DihedralGroup 4)) ≃* (DihedralGroup 4)) := by
  sorry 
\end{lstlisting}

\paragraph{Exercises with diagrams}

Some exercises include diagrams, such as subgroup lattices or commutative diagrams. For example, Exercise 14 of Chapter 2.5 concerns the \emph{modular group} of order $16$, defined by the presentation $M := \langle u, v \mid u^2 = v^8 = 1,\ vu = uv^5 \rangle$.
In particular, it asks the reader to show that the lattice of subgroups of $M$ is the same as the lattice of subgroups of $\mathbb{Z}/2\mathbb{Z} \times \mathbb{Z}/8\mathbb{Z}$. The \lstinline{Subgroup} type in \texttt{Mathlib} is already an instance of \lstinline{CompleteLattice}, so this can be formalized as follows:
\begin{lstlisting}[language=Lean, style=leanplain]
inductive MGen | u | v deriving DecidableEq, Repr

open FreeGroup in
def M := PresentedGroup {
  (of MGen.u)^2,
  (of MGen.v)^8,
  (of MGen.u) * (of MGen.v)^5 * (of MGen.u)^(-1 : ℤ) * (of MGen.v)^(-1 : ℤ)
}

deriving instance Group for M

theorem DF_2_5_14_4 : Nonempty ((Subgroup M) ≃o AddSubgroup ((ZMod 2) × (ZMod 8))) := by
  sorry
\end{lstlisting}
Here \lstinline{≃o} denotes \lstinline{OrderIso}, an order-preserving isomorphism.

\begin{figure}[htbp]
\centering
\begin{adjustbox}{max width=\textwidth}
$
\begin{tikzcd}[
  row sep=4.5em,
  column sep=2.2em,
  cells={nodes={inner sep=1pt}},
  every arrow/.append style={line width=.7pt}
]
&&&&& G
  \arrow[no head, from=1-6, to=2-4]
  \arrow[no head, from=1-6, to=2-6]
  \arrow[no head, from=1-6, to=2-8]
  \arrow[no head, from=2-4, to=3-2]
  \arrow[no head, from=2-4, to=3-4]
  \arrow[no head, from=2-4, to=3-6]
  \arrow[no head, from=2-6, to=3-6]
  \arrow[no head, from=2-8, to=3-6]
  \arrow[no head, from=2-8, to=3-8]
  \arrow[no head, from=2-8, to=3-10]
  \arrow[no head, from=3-2, to=4-1]
  \arrow[no head, from=3-2, to=4-3]
  \arrow[no head, from=3-2, to=4-6]
  \arrow[no head, from=3-4, to=4-4]
  \arrow[no head, from=3-4, to=4-5]
  \arrow[no head, from=3-4, to=4-6]
  \arrow[no head, from=3-6, to=4-6]
  \arrow[no head, from=3-8, to=4-6]
  \arrow[no head, from=3-10, to=4-6]
  \arrow[no head, from=4-1, to=5-6]
  \arrow[no head, from=4-3, to=5-6]
  \arrow[no head, from=4-4, to=5-6]
  \arrow[no head, from=4-5, to=5-6]
  \arrow[no head, from=4-6, to=5-6]
&&&& \\
&&& {G_0 = \langle \sigma^2,\tau\rangle} && {G_1 = \langle \sigma\rangle}
  && {G_2 = \langle \sigma^2,\tau\sigma\rangle} && \\
& {A_0} && {H_0 = \langle \sigma^4,\tau\rangle}
  && {A_1} && {A_2}
  && {A_3} \\
{H_1 = \langle \tau\sigma^2\rangle} && {A_4}
  & {A_5} & {H_2 = \langle \tau\rangle} & {H_3 = \langle \sigma^4\rangle}
  &&&& \\
&&&&& 1 &&&&
\end{tikzcd}
$
\end{adjustbox}
\caption{Subgroup lattice for Exercise 11, Chapter 2.5}
\label{fig:subgroup-lattice}
\end{figure}
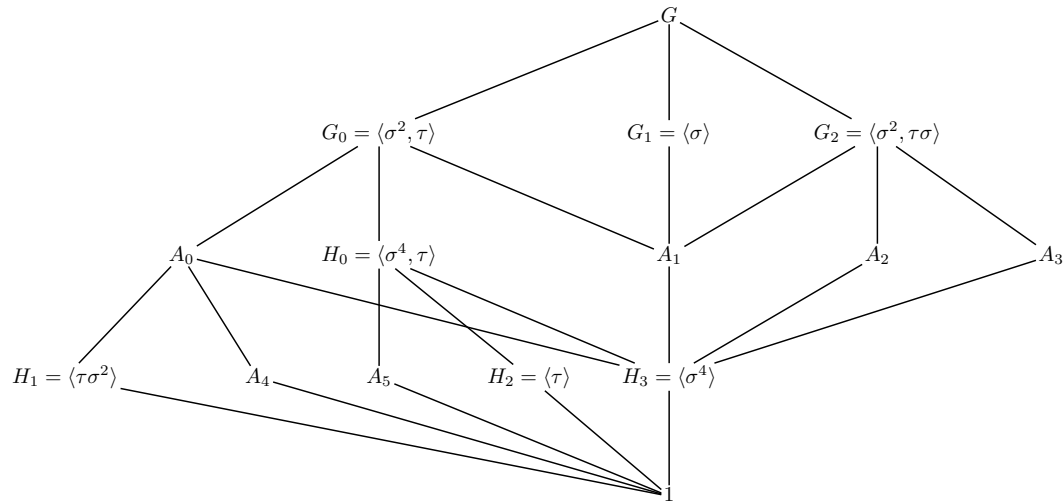

A similar but more complicated example is Exercise 11 of the same chapter, which concerns the subgroup lattice of the \emph{quasidihedral group} of order $16$,
\[
G = QD = \langle \sigma, \tau \mid \sigma^8 = \tau^2 = 1,\ \sigma \tau = \tau \sigma^3 \rangle.
\]
The exercise provides a subgroup lattice diagram with six missing vertices, shown in Figure~\ref{fig:subgroup-lattice}, and asks the reader to identify suitable subgroups $A_0, A_1, \dots, A_5$, each generated by at most two elements.

As with other problems that ask for explicit answers, we define \lstinline{DF_2_5_11_ans} using \lstinline{sorry}s. These answers have type \lstinline{Set QD}, and \lstinline{DF_2_5_11_groups} is defined to be the corresponding collection of subgroups of $QD$ generated by the sets in \lstinline{DF_2_5_11_ans}. In other words, a solution must provide generating sets for the missing groups $A_0, \dots, A_5$. A complete formalization is given below:
\begin{lstlisting}[language=Lean, style=leanplain]
inductive QDGen | σ | τ deriving DecidableEq, Repr

open FreeGroup in
def QD := PresentedGroup {
  (of QDGen.σ)^8,
  (of QDGen.τ)^2,
  (of QDGen.σ) * (of QDGen.τ) * (of QDGen.σ)^(-3 : ℤ) * (of QDGen.τ)^(-1 : ℤ)
}

deriving instance Group for QD


abbrev QD.of (x : QDGen) : QD := PresentedGroup.of x

abbrev σ : QD := PresentedGroup.of QDGen.σ
abbrev τ : QD := PresentedGroup.of QDGen.τ

def DF_2_5_11_ans : Fin 6 → Set QD
  | 0 => sorry
  | 1 => sorry
  | 2 => sorry
  | 3 => sorry
  | 4 => sorry
  | 5 => sorry

def DF_2_5_11_groups : Fin 6 → Subgroup QD := fun i ↦ Subgroup.closure (DF_2_5_11_ans i)

def G₀ := Subgroup.closure {σ ^ 2, τ}

def G₁ := Subgroup.closure {σ}

def G₂ := Subgroup.closure {σ ^ 2, τ * σ}

def H₀ := Subgroup.closure {σ ^ 4, τ}

def H₁ := Subgroup.closure {τ * σ ^ 2}

def H₂ := Subgroup.closure {τ}

def H₃ := Subgroup.closure {σ ^ 4}

theorem DF_2_5_11_1 : ∀ i : Fin 6, Finite (DF_2_5_11_ans i) ∧ (DF_2_5_11_ans i).ncard ≤ 2 := by
  sorry

theorem DF_2_5_11_2 :
    DF_2_5_11_groups 0 < G₀ ∧
    H₁ < DF_2_5_11_groups 0 ∧
    H₃ < DF_2_5_11_groups 0 ∧
    DF_2_5_11_groups 1 < G₀ ∧
    DF_2_5_11_groups 1 < G₁ ∧
    DF_2_5_11_groups 1 < G₂ ∧
    H₃ < DF_2_5_11_groups 1 ∧
    DF_2_5_11_groups 2 < G₂ ∧
    H₃ < DF_2_5_11_groups 2 ∧
    DF_2_5_11_groups 3 < G₂ ∧
    H₃ < DF_2_5_11_groups 3 ∧
    DF_2_5_11_groups 2 ≠ DF_2_5_11_groups 3 ∧
    DF_2_5_11_groups 4 < DF_2_5_11_groups 0 ∧
    ⊥ < DF_2_5_11_groups 4 ∧
    DF_2_5_11_groups 5 < H₀ ∧
    ⊥ < DF_2_5_11_groups 5 := by
  sorry
\end{lstlisting}
The theorem \lstinline{DF_2_5_11_1} asserts that the proposed generating sets have at most two generators, while \lstinline{DF_2_5_11_2} asserts the inclusion relations among the subgroups described in Figure~\ref{fig:subgroup-lattice}. Since $A_2$ and $A_3$ are both subgroups of $G_2$ and both contain $H_3$, we also require a proof that the two proposed subgroups are distinct, namely \lstinline{DF_2_5_11_groups 2 ≠ DF_2_5_11_groups 3}.

\paragraph{Exercises beyond current autoformalization}

Some exercises lie outside what current autoformalization systems can handle. 
Despite running a wide range of systems, from frontier single-shot models to 
recent open-weight models and a compiler-grounded agent loop (Section~\ref{sec:elaboration}),
producing a usable Lean file without human curation remained out of reach
for a non-trivial fraction of exercises, and the difficulty grows quickly 
as the underlying mathematics becomes more advanced.
Exercise 2 of Chapter 16.2 illustrates that even short,
self-contained statements can fall into this category:
\begin{textbox}
  Suppose $R$ is a discrete valuation ring with unique maximal ideal $M$ and quotient $F = R/M$. For any $n \ge 0$, show that $M^n / M^{n+1}$ is a vector space over $F$ and that $\dim_F(M^n / M^{n+1}) = 1$.
\end{textbox}
The exercise does not require any answer to be supplied, 
and every ingredient needed for its formalization 
(discrete valuation rings, the residue field, and the quotient $M^n / M^{n+1}$ as an $F$-module)
is already available in \texttt{Mathlib}.
Nevertheless, all of our single-shot models failed to elaborate on this exercise. 
Codex did produce a file that elaborates, but the resulting statement encoded 
neither the defining property of a discrete valuation ring nor the correct 
quantifier structure of the informal problem, so the formalization does not 
match the original.

This points to a more general limitation: when an autoformalizer does not 
understand the mathematics behind a statement, it tends to copy the words 
rather than produce a faithful formalization, even when the problem looks 
routine and the needed ingredients are already in \texttt{Mathlib}.
We expect this gap to persist, if not widen, as topics become more advanced. 
Human curation is therefore essential, not an optional safety net: 
without it, even a model that elaborates well can silently produce a wrong 
statement on an exercise that looks easy at first glance. 
Our manual formalization is as follows:
\begin{lstlisting}[language=Lean, style=leanplain]
theorem DF_16_2_2'
    {R : Type*} [CommRing R] [IsDomain R] [IsDiscreteValuationRing R] (n : ℕ) :
    let M := IsLocalRing.maximalIdeal R
    let F := IsLocalRing.ResidueField R
    let V := (M ^ n : Ideal R) ⧸ (Submodule.comap (M^n).subtype (M^(n+1)))
    ∃ _ : Module F V, Module.finrank F V = 1 := by
  sorry
\end{lstlisting}

\subsection{Compute resources}
\label{app:compute}

All open-weight autoformalization runs were performed on a single NVIDIA H100 NVL (96\,GB) or H200 NVL (144\,GB) GPU on an internal academic HPC cluster; no multi-GPU or distributed inference was used. Models were served through Ollama with each model's default sampling parameters, and we performed no training, fine-tuning, or weight modification. Lean elaboration of the generated files was performed post-hoc on a CPU host using \texttt{leanprover/lean4:v4.29.1} and the \texttt{Mathlib} revision \lstinline{5e932f9}, and is not GPU-bound.

\begin{table}[t]
  \centering
  \caption{Per-run wall-clock time and Ollama-reported token counts for the four open-weight single-shot runs in Section~\ref{sec:elaboration}. All runs use Q4\_K\_M quantization except Goedel-Formalizer-V2, which is only available at Q8\_0 from its public Ollama upload.}
  \label{tab:compute}
  \vspace{.2cm}
  \begin{tabular}{lcrr}
    \toprule
    Run & GPU & Wall-clock & Prompt / Output tokens (M) \\
    \midrule
    Gemma4-31B (think)             & H200 NVL       & 43.9\,h  & 0.67 / 7.56  \\
    Goedel-Formalizer-V2-32B       & H200 NVL       & 10.3\,h  & 0.65 / 1.83  \\
    Qwen3.6-35B-A3B                & H100/H200 NVL  & 64.4\,h  & 0.67 / 17.54 \\
    DeepSeek-R1-Distill-Qwen-32B   & H200 NVL       & 46.6\,h  & 0.64 / 5.26  \\
    \midrule
    Subtotal                       &                & 165.2\,h & 2.63 / 32.19 \\
    \bottomrule
  \end{tabular}
\end{table}

The closed-source GPT-5, Codex (agent loop with GPT-5 backbone), and Claude Opus~4.7 (LLM judge, Section~\ref{subsec:semantic-check}) were accessed through provider APIs; their compute is provider-side and is not included in the cluster totals above. Exploratory open-weight runs that did not appear in our final comparison consumed approximately 80 additional GPU-hours on the same hardware.


\end{document}